\documentstyle[12pt, epsf]{article}

\renewcommand{\thefootnote}{\fnsymbol{footnote}}

\newcommand{\non}{\nonumber \\*}

\newcommand{\eq}[1]{Eq.~(\ref{#1})}
\def \ov {\over }
\def\bea{\begin{eqnarray}}
\def\eea{\end{eqnarray}}

\def \de{\partial}
\def \x{{\bf x}}
\def\y{{\bf y}}
\def\k{{\bf k}}
\def\LB{\left(}
\def\RB{\right)}
\def\be{\begin{equation}}
\def\ee{\end{equation}}
\def\la{\label}
\def \bi{\bibitem}
\def \Tr {{\rm tr}}
\def\O{{\cal O}}
\def\C{{\cal C}}
\def\LA{\langle}
\def\RA{\rangle}
\def\X{{\bar X}}

\font\mybbb=msbm10 at 8pt

\def\bbb#1{\hbox{\mybbb#1}}

\def\pRe{\bbb{R}}

\begin{document}

\setcounter{page}{1}
\renewcommand{\thefootnote}{\arabic{footnote}}
\setcounter{footnote}{0}

\begin{titlepage}
\begin{flushright}
ITP-SB-99-29\\
\end{flushright}
\vspace{.5cm}

\begin{center}
{\LARGE  A note on correlation functions in ${\rm AdS}_5/{\rm SYM}_4$ 
correspondence
on the Coulomb branch}\\
\vspace{1.1cm}
{\large Iouri Chepelev${}$\footnote{
E-mail: chepelev@insti.physics.sunysb.edu } 
and  Radu Roiban${}$\footnote{ \ E-mail: roiban@insti.physics.sunysb.edu } }\\
\vspace{18pt}

{\it Institute for Theoretical Physics}

{\it SUNY at Stony Brook}

{\it  NY11794-3840, USA}
\\
\end{center}
\vskip 0.6 cm

\begin{abstract}
We compute certain two-point functions in $D=4$, ${\cal N}=4$, SU(N) SYM theory
on the Coulomb branch  using  SUGRA/SYM duality and find an infinite set
of first order poles at masses of  order $({\rm Higgs~scale})/(g_{YM} \sqrt{N})$.

\end{abstract}

\end{titlepage}

\setcounter{footnote}{0}
\noindent
{\bf 1.}~Apart from being a universal way of looking at the gravity/gauge theory
 relation, 
Maldacena's supergravity(SUGRA)/super Yang-Mills(SYM) duality 
\cite{malda, IMSY} provides  a 
useful tool for the study  of strongly coupled gauge 
theories.\footnote{For a recent review of SUGRA/SYM duality see ref.\cite{AGM}.}
 The precise  rules for relating SUGRA and SYM observables in the context of 
${\rm AdS}_5/{\rm CFT}_4$ duality of ref.\cite{malda}
were given in refs.\cite{GKP,witten}. 
They consist of
(A) the dictionary between gauge-invariant SYM operators and 
   SUGRA fields, and
(B) the precise relation between the generating functional of
correlators of ${\cal N}=4$ ${\rm CFT}_4$ and the Type IIB  SUGRA 
action $S(\phi^i)$ evaluated on the solutions $\phi^i$ of the
classical equations satisfying the boundary conditions $\left. \phi^i \right|_{\de 
AdS}\sim \phi_0^i$, where $\phi_0^i$ are the sources for the operators
of ${\rm CFT}_4$. 
\be
\left\langle {\rm exp} \int_{S^4} \phi_0^i {\cal O}_i 
\right\rangle_{CFT}={\rm exp}(-S(\phi^i))  \, .
\la{relation}
\ee

Rules A and B were used for the calculation of 
two, three and four point functions of strongly coupled conformal
${\cal N}=4$ SYM in refs.[4--11].  
Given that in the conformal case one can use SUGRA/SYM duality to 
calculate correlation functions of ${\rm SYM}_4$, one naturally wonders if
rules A and B can be modified to incorporate 
${\rm SYM}_4$ on the
Coulomb branch. The ${\rm AdS}_5/{\rm SYM}_4$ correspondence on the Coulomb branch
was first discussed in ref.\cite{malda}, where the
relevance of the multi-center $D3$-brane SUGRA solution to 
${\rm SUGRA}/{\rm SYM}_4$ duality
 was first mentioned. 
Various aspects of this correspondence have since 
been discussed in ref.\cite{coulomb}, but it
stood on  much weaker grounds compared to the correspondence 
in the conformal
case. Recently, Klebanov and Witten \cite{KW}  
gave arguments, partly based on 
an earlier work  ref.\cite{balasub}, 
in favor of AdS/SYM correspondence
on the Coulomb branch.

Let us recall a  generalization of rule B
proposed in ref.\cite{balasub}. There it was pointed out that 
$
{\delta (-S(\phi))\ov \delta \phi_0(\x)}=
\LA {\cal O}(x)\RA_{\phi_0}
$ is
the expectation value of the operator ${\cal O}(x)$ in the 
presence of 
the boundary source $\phi_0$. In the context of \eq{relation}, the bulk
solution $\phi$ to the classical field equations is  completely
determined by the boundary value $\phi_0$ and the requirement of regularity in the bulk. 
This uniqueness fails if one admits singular fields corresponding
to sources in the bulk. The generalization of \eq{relation} proposed
in ref.\cite{balasub} consists in introducing sources in the bulk. 
The SYM one-point function $\LA\Tr F^2(\x)\RA$ in the
instanton background was computed in ref.\cite{balasub} by considering 
the response of bulk SUGRA action in  the D-instanton 
background to the change in boundary data.
In the same sense as the relevant SUGRA background to 
consider in the case of SYM in the instanton background is
AdS D-instanton background, multi-center D3-brane background is relevant
in the case of  SYM on the Coulomb branch \cite{KW}.

\noindent
{\bf 2.}~In the present work we  extend  rules A and B 
to the case of ${\rm SYM}_4$ with the Higgs vev $\X$ turned on, and
apply the modified rules to the specific case of spherically
symmetric distribution of eigenvalues of $\X$. 
Let $\O_i=\O_i[F,X]$ be gauge invariant SYM quantum operators.
We are interested in the connected Green's
functions 
\be
\LA \O_1(\x_1) \dots \O_n(\x_n)\RA_{\X}={1\over Z}\int {\cal D}A 
{\cal D}X {\cal D}\psi \, 
e^{-S[A,X+\X, \psi]} \O_1(\x_1) 
\dots \O_n(\x_n)\Bigg|_{{\rm conn.}}  \, .
\ee
Note that $\LA \O_i \RA_{\bar X}=0$.

Consider the 10D Type IIB SUGRA action 
$S_{10}(\Phi_i)$ in 
the multi-center D3-brane background:
$$
ds^2=H^{-1/2}(dt^2+dx_1^2+dx_2^2+dx_3^2)+H^{1/2}\sum_{j=1}^6 dy_j^2  
$$
\be
H({\vec y})=Q\int d^6 y' {\rho({\vec y'}) \over |{\vec y}-{\vec y'}|^4}\,\, \,
 ,
\la{back}
\ee
where $\rho$ is the distribution function of D3-branes normalized as 
$\int d^6 y \rho({\vec y}) = N$	and Q is the charge of a 
single D3-brane. 
Now substitute the expansions of
the fields ${\Phi_i}$ in spherical harmonics
  \be
\Phi_i(\x, {\vec y})=\sum \phi^I_i(\x,r) Y^I(\Omega_5) \,,~~~~~
r\equiv |{\vec y}|
\ee
into $S_{10}(\Phi_i)$ and integrate
over the sphere ${\rm S}^5$. We end up with a five dimensional
action $S_5=S_5(\phi_i^I)$ in some effective background which is
asymptotically ${\rm AdS}_5$.\footnote{In general there will  be
no consistent truncation of $S_5$ to lowest KK modes $\phi_i^{I=0}$. There is
actually no  fundamental reason for the existence of such a truncation in the 
generic
case. The large N properties of ${\rm SYM}_4$ are encoded in  10D Type IIB 
SUGRA on appropriate backgrounds and not in 5D truncations thereof.}     
In the single-center D3-brane case $S_5$
coincides with the five dimensional gauged SUGRA action on
${\rm AdS}_5$ background with the infinite tower of Kaluza-Klein fields 
included. 
In the conformal case  rule A was formulated  
by 
matching  the spectrum of SUGRA fields on the ${\rm AdS}_5\times
{\rm S}^5$ background \cite{peter} with the conformal dimensions of SYM 
operators \cite{witten}. A natural extension of rule A to  our case consists in  matching
the  spectrum of SUGRA fields in the infrared with the dimensions of
the SYM operators in the ultraviolet. The modification of rule B reads as 
follows  
\be
\left\langle {\rm exp} \int_{\pRe^4} \phi_0^{iI} {\cal O}_{iI} 
\right\rangle_{{\bar X}}={\rm exp}(-S_5(\phi_i^I)) \,  .
\la{relation1}
\ee
Since asymptotically the geometry is ${\rm AdS}_5$, we impose  
the same boundary conditions for the bulk SUGRA fields $\phi_i^I$
as in the conformal case. 

\noindent
{\bf 3.}~Let us apply the modified rules  to the computation of the two-point functions
of the operators $\Tr (F^2 X^I)$,
where $X^I=X^{i_1}X^{i_2} \cdots X^{i_l} \C_{i_1 i_2 \cdots i_l}^I$  \cite{seiberg}.
These operators correspond to the
Kaluza-Klein harmonics $\phi^I$ of the 10D dilaton. Consider the
 dilaton kinetic term in $S_{10}$ and Kaluza-Klein reduce it to 5D as
described earlier. 
\bea
&&~~~~~~~~~\int d^{10}x (H \Phi \nabla_{||}^2 \Phi + \Phi \nabla_{\perp}^2 
\Phi )\rightarrow  \non
&&\rightarrow 
\int dx_{||}^4 dr r^5 \left[\phi 
( \LB \int d\Omega_5 H\RB \nabla_{||}^2  + V_5 {1\over r^5} 
\partial_r r^5 \partial_r ) \phi \right. \non
&& +\LB \int d\Omega_5 H Y^I \RB (\phi \nabla^2_{||} \phi^I+\phi^I
\nabla_{||}^2 \phi ) +
\LB \int d\Omega_5 H Y^I Y^J \RB \phi^I  \nabla^2_{||} \phi^J
\non
&&+
\LB \int d\Omega_5 Y^I Y^J\RB \phi^I   {1\over r^5} 
\partial_r r^5 \partial_r  \phi^J +	 \left.
{1\over r^2} \LB    \int d\Omega_5 Y^I \nabla^2_{S^5} Y^J\RB	 
\phi^I \phi^J \right]  \, .
\la{harmon}
\eea
where $\phi=\phi^{I=0}$ and $V_5=\int d \Omega_5$.
The form of
the action \eq{harmon} suggests that the correlation function
$\LA \Tr F^2 \Tr (F^2 X^I) \RA_{{\bar X}}$ is non-vanishing for  generic
Higgs vev ${\bar X}$. However, in the spherically symmetric case 
$\rho=\rho(|{\vec y}|)$ the coupling
between $\phi$ and $\phi^I$ vanishes, implying that this correlator
vanishes. 

Let us show how this happens in SYM. The distribution
of D3-branes in SUGRA corresponds to the distribution of
Higgs vev $\X^i$ in SYM. 
The general form of the correlator $\LA \Tr (F^2 X^I)(\x) 
\Tr (F^2 X^J)(\y)\RA_{\X}$ compatible with gauge and R symmetries is 
$$
f_1\LB|\x-\y|, 
\Tr (\X^J \X_J), \Tr \X^J \Tr \X_J \RB \Tr (\X^I \X^J)
$$
\be
+ f_2\LB|\x-\y|, 
\Tr (\X^J \X_J), \Tr \X^J \Tr \X_J\RB \Tr \X^I \Tr \X^J  \, ,
\ee
where $f_1$ and $f_2$ are some arbitrary functions. In the spherically 
symmetric case only the first term survives and it is 
proportional to $\delta^{I J}$,  in agreement with SUGRA.

Now consider the most general finite, spherically symmetric 
distribution of D3 branes. From \eq{back} we see that the 
harmonic function becomes
\be
H(\vec y)=\begin{array}{r}
\left\{ \begin{array}{rr} 
f(|y|)\,\,\,\,\,\,\,\,|y|\le r_0 \\
\frac{NQ}{|y|^4}\,\,\,\,\,\,\,\,r_0\le |y|
       \end{array} \right. \, .
\end{array}
\ee
As we argued in \eq{harmon}, in this case the KK harmonics are decoupled.
In consequence, we can study them separately. The equations of motion for 
an arbitrary mode $\phi^I$ in this backgrond are:
$$
{1\ov r^5}\de_r (r^5 \de_r \phi^I)+\frac{q(I)}{r^2}\phi^I-k^2 f(r)\phi^I=0 \, ~~~r\le r_0
$$
\be
{1\ov r^5}\de_r (r^5 \de_r \phi^I)+\frac{q(I)}{r^2}\phi^I-{k^2 NQ\ov r^4}\phi^I=0 \, ~~~r_0\le r  \,  ,
\la{eigen}
\ee
where $q(I)=-l(l+4)$ is the eigenvalue asociated to spherical harmonics $Y^I$.
Let $\chi$ and $\Psi$ be the two solutions of the equation for 
$r\le r_0$ and assume that $\chi$ is well behaved at zero while $\Psi$
is well behaved at infinity. Then, matching the solutions of \eq{eigen}
at $r=r_0$ we have for the solutions in the interval
$r\in [0,\infty )$:

\noindent
$\bullet$ Solution well behaved at $r=\infty$ ($\xi=\frac{r_0}{r}=0$):
\be
\psi^I_1(\xi)=\left\{ 
\begin{array}{cl}
\Psi(\xi,I)+ \gamma^I(\kappa)\chi(\xi,I) & r\le r_0\\
\xi^2 I_{l+2} \LB {\kappa\xi}\RB & r>r_0
\la{sol1}
\end{array}
\right.  \, .
\ee
Here $\kappa^2={k^2 N Q\ov r_0^2}$.

\noindent
$\bullet$ Solution well behaved at $r=0$ ($\xi=\frac{r_0}{r}=\infty$):
\be
\psi^I_2(\xi)=\left\{ 
\begin{array}{cl}
\chi(\xi,I) & r\le r_0 \\
\xi^2 K_{l+2}\LB {\kappa\xi}\RB + \beta^I(\kappa) \xi^2 I_{l+2}\LB {\kappa\xi} 
\RB & r>r_0
\la{sol2}
\end{array}
\right. \,  .
\ee
Let us give two examples. 
For the case of the spherical shell distribution
\be
\rho(|y|)={N\ov V_5 r_0^5} \delta (r-r_0)
\ee
we have $f(|y|)={QN\ov r_0^4}$ and 
\be
\chi(\xi,I)= \xi^2 I_{\nu}\LB{\kappa\ov\xi}\RB,~~~\nu=2+l \, .
\ee
For the case of uniform distribution
\be
\rho(|y|)=\begin{array}{r}
\left\{ \begin{array}{rr} 
\frac{6N}{V_5r_0^6}\,\,\,\,\,\,\,\,|y|\le r_0 \\
0\,\,\,\,\,\,\,\,r_0\le |y|
       \end{array} \right.
\end{array}
\la{unif}
\ee
we have
$f(|y|)=\frac{QN}{r_0^6}(3r_0^2-2r^2)$ and 
\be
\chi(\xi,I)=\xi^{-l} \,\,{\sl e}^{i {\kappa\ov\sqrt{2}} 
\xi^{-2}}{ }_1F_1[{3\ov 2}+{l\ov 2}+
i{3\sqrt{2}\kappa\ov 8},l+3,-\sqrt{2}i\kappa \xi^{-2}]  \, .
\ee

Since we are interested in the limit $r\rightarrow\infty$
($\xi\rightarrow 0$), only $\beta(\kappa)$ is relevant. It reads
\be
\beta^I(\kappa)=-\frac{\chi(\xi,I)\partial_\xi(\xi^2 K_{l+2} \LB {\kappa\xi}\RB )-
\partial_\xi\chi(\xi,I)(\xi^2 K_{l+2}\LB {\kappa\xi}\RB)}
{\chi(\xi,I)
\partial_\xi(\xi^2 I_{l+2} \LB {\kappa\xi}\RB )-
\partial_\xi\chi(\xi,I)(\xi^2 I_{l+2}\LB {\kappa\xi}\RB)}\Bigg|_{\xi=1}  \, .
\ee
With this solution the scalar Green's function has the following expression
\be
G^I_{\epsilon}(x,y)=G^I_0(x,y)+\int {d^4k\ov (2\pi)^4}e^{-i \k (\x-\y)} \psi^I_2(x_0)
\psi^I_2(y_0){\psi^I_1(\epsilon)\ov \psi^I_2(\epsilon)}\, ,
\ee
where
\be
G^I_0(x,y)=-\int {d^4k\ov (2\pi)^4} e^{-i \k (\x-\y)}\left\{
\begin{array}{cl}
\psi^I_1(x_0)\psi^I_2(y_0)& x_0<y_0 \\
\psi^I_1(y_0)\psi^I_2(x_0)& x_0>y_0  
\end{array}
\right. \, .
\ee

Using the fact that the action on this solution is
\be 
S[\phi]=-{1\ov 2} \lim_{\epsilon\rightarrow 0}\int d^4x \epsilon^{-3} \left. \phi^I \de_0 \phi^I \right|_{x_0=\epsilon}
\ee
with
\bea
\phi^I(\x,x_0)&=&\int d^4 y \left. (y_0^{-3}{\de \ov \de y_0} G^I_{\epsilon}(x,y))\right|_{
y_0=\epsilon} \phi_0^I (\y) \epsilon^{-l}\non
&=&
\int d^4 y \phi_0^I (\y)\epsilon^{-l} \int {d^4k\ov (2\pi)^4} e^{-i \k (\x-\y)} {\psi_2^I(x_0)\ov \psi_2^I(\epsilon)}  \, \, \, ,
\eea
we get the two-point function
\bea
\LA\O^I(\x) \O^J(\y) \RA= -\delta^{IJ} \int {d^4 k\ov (2\pi)^4}\!\!\!\!\!\!&{ }&\!\!\!\!\!\!
e^{-i \k (\x-\y)}\LB {2^{1-2\nu} \Gamma(1-\nu)\ov \Gamma(\nu)} 
(k \, \sqrt{QN})^{2\nu}-\right. \nonumber\\
{ }~~~~~~~~~~~~~~~~&-&\!\!\left.{2^{2(1-\nu)}\nu
\ov \Gamma(\nu)\Gamma(\nu +1)} \beta^I(\kappa) (k \, \sqrt{QN})^{2\nu} \RB \, \, ,
\la{tpf}
\eea

\noindent
where $\nu=2+l$,
and after performing the $k$ integral we find
\bea
&&{\LA \O^I(\x) \O^J(\y) \RA\ov (QN)^{\nu}}= \non
&&=\delta^{IJ}{2\nu^2 (1+\nu) \ov \pi^2 |\x-\y|^{4+2\nu}}(1+{2^{-(2\nu+1)}\ov \Gamma (\nu+1) \Gamma (\nu+2)} 
\int_0^{\infty} dk k^6 J_1(k) \beta^I \LB {k{\sqrt{QN}\ov r_0 |\x-\y|}}\RB ) \,  .
\eea
In the extreme UV limit,  $|\x-\y|r_0/\sqrt{QN} \ll 1$,  we have $\beta^I\rightarrow 0$ 
and we recover CFT correlators. Note that CFT behaviour is valid down to a much
lower energy scale ${r_0\ov g_{YM}\sqrt{N}}$ compared to the Higgs scale $r_0$.

\noindent
{\bf 4.}~Performing the Wick rotation $k\rightarrow -i\, k$ to Minkowski space one 
opens the possibility of studying at 
least part of the spectrum of SYM theory on the Coulomb branch. 
Recalling the transformations of Bessel functions
$
I_\nu(-i\,z)=e^{-i\pi{\nu\ov 2}}J_\nu(z)$,
$K_\nu(-i\,z)={i\pi\ov 2} e^{+i\pi{\nu\ov 2}}\LB J_{\nu}(z)+iY_\nu(z)\RB$
we find
\be
\beta^I(\kappa)\rightarrow (-1)^l{i\pi\ov 2}+(-1)^l{\pi\ov 2} \frac{\tilde \chi(\xi,I)\partial_\xi(\xi^2 Y_{l+2} \LB {\kappa
\xi}\RB )-
\partial_\xi\tilde\chi(\xi,I)(\xi^2 Y_{l+2}\LB {\kappa\xi}\RB)}
{\tilde\chi(\xi,I)
\partial_\xi(\xi^2 J_{l+2} \LB {\kappa\xi}\RB )-
\partial_\xi\tilde\chi(\xi,I)(\xi^2 J_{l+2}\LB {\kappa\xi}\RB)}\Bigg|_{\xi=1}
\, \, \, ,
\la{wick}
\ee
where $\tilde\chi$ is the Wick rotated solution of the wave equation for 
$r\le r_0$ 
which is well-behaved at $r=0$. The first term in \eq{wick} cancels against a 
similar 
term coming from the conformal piece of the two-point function (first 
term in \eq{tpf}). At this point we can read of the singularity 
structure of 
the two-point function (see figures 1 and 2). The poles are given by the 
solutions of the equation
\be
\tilde\chi(\xi,I)
\partial_\xi(\xi^2 J_{l+2} \LB {\kappa\xi}\RB )-
\partial_\xi\tilde\chi(\xi,I)(\xi^2 J_{l+2}\LB {\kappa\xi}\RB)\Bigg|_{\xi=1}=0
\, .
\label{poles}
\ee
The solutions of \eq{poles} which are first order zeroes correspond to 
states in SYM which 
are color singlets, have the right quantum numbers to couple to $\O^I=\Tr (F^2 X^I)$ and have 
masses of  order ${r_0\ov g_{YM}\sqrt{N}}$. 
\begin{figure}[hbtp]
\begin{center}
\mbox{\epsfxsize=9truecm
\epsffile{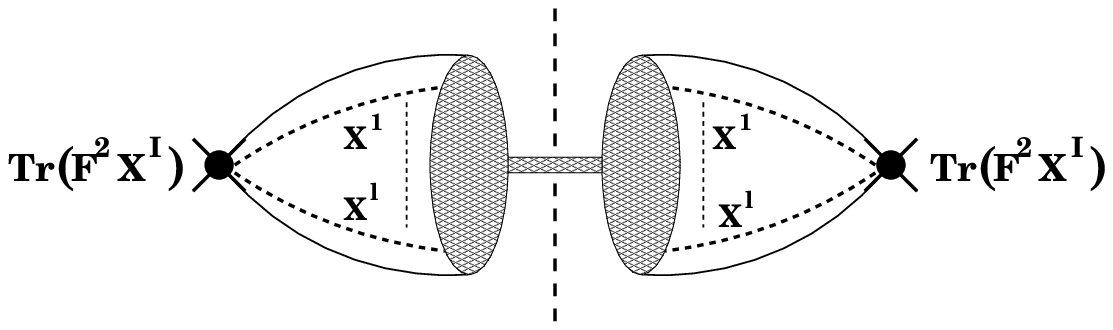}}
\end{center}
\centerline{{\bf Fig. 1.}}
\end{figure}
In the large $N$ limit these states are stable. However, for finite N they 
become unstable
against decay into photons. The strength of the coupling can be read from 
the two-point function.

The same set of states can be obtained in the approach of 
ref.\cite{terning}. One starts with the wave equation for the mode $\phi^I$
and solves it as an eigenvalue equation subject to  suitable boundary 
conditions required by normalizability. In our case, the solutions are
those given in \eq{sol1} 
and \eq{sol2}. Imposing normalizability of the solution at $r=0$ singles out 
the solution in \eq{sol2}, while normalizability at infinity
implies that the coefficient of $\xi^2K_{l+2}(\kappa\xi)$ (the denominator of 
$\beta^I(\kappa)$) is zero. This is the Wick-rotated form of 
\eq{poles} giving therefore the same spectrum.

For the uniform distribution discussed before, a plot of the Wick-rotated 
function $\beta$ for $l=0$ looks as in figure 2. 
\begin{figure}[hbtp]
\begin{center}
\mbox{\epsfxsize=9truecm
\epsffile{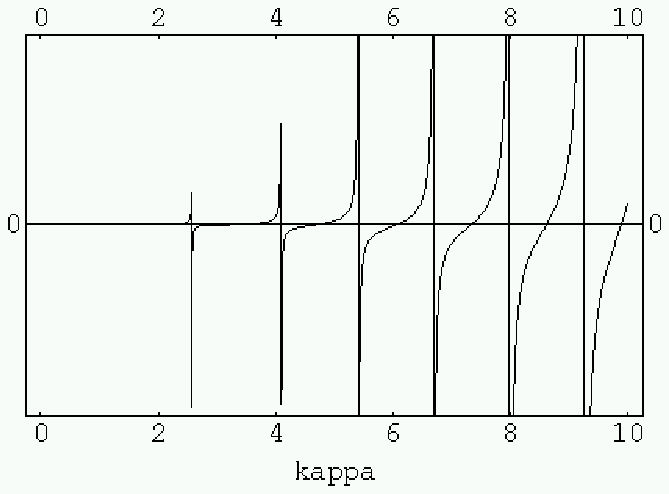}}
\end{center}
\centerline{{\bf Fig. 2}: Plot of $\beta^{I=0}(\kappa)$ for the distribution \eq{unif}.}
\end{figure}

\noindent
Using the explicit solution of the wave equation and/or numerics, one can show that all
the poles are simple poles and thus correspond to physical states.
Numerical analysis shows that the poles move towards higher masses as one 
increases $I$. Similar behaviour arises for the spherical shell distribution.

\noindent
{\bf 5.}~One may wonder whether the poles of the two-point function we
found are artifacts of  continuous distribution of D3 branes. It would 
be very interesting to compute some two-point functions for
the case of the SYM with the gauge group $U(2N)$ broken to $U(N)\times
U(N)$ using the two-center D3 brane solution. 
For this purpose one has to solve 10D wave equation following from
the 10D action \eq{harmon} on the two-center D3 brane background.
We expect that the two-point functions will have poles in this case as well.

\pagebreak

\centerline {\ \bf Acknowledgments}
We are grateful to H.~Nastase and M.~Ro\v{c}ek for useful discussions. We
would also like to thank I.~Klebanov, J.~Maldacena, P.~van Nieuwenhuizen,
H.~Ooguri, L.~Rastelli, D.~Vaman and T.T.~Wu for stimulating discussions.

\bigskip

\noindent
{\bf Note Added} 

As this work was being completed, there appeared refs.\cite{fgpw,sfet} 
which discuss some other distributions of D3 branes and give conceptually
different interpretation of AdS/SYM correspondence on the Coulomb branch. 
In ref.\cite{fgpw} 
the extreme smallness of masses at the poles of the two-point functions
was interpreted as  an artifact of SUGRA 
approximation. The argument
was based on the fact  that the curvatures of the geometries which were 
considered 
become large close to the brane distribution and therefore, at low energies,
supergravity is not reliable. Our uniformly distributed branes solution
does not share this feature, but we still find masses of the same order
of magnitude, suggesting that the unnatural smallness of masses is not an 
artifact of SUGRA approximation.

\end{document}